\documentclass[aps,pra,twocolumn,groupedaddress,longbibliography]{revtex4-1}
\usepackage{graphicx}% Include figure files
\usepackage{dcolumn}% Align table columns on decimal point

\begin{document}
	
	%\preprint{}
	
	\title{Preparation of circular Rydberg states in helium with $n \geq 70$ using a modified version of the crossed-fields method}
	
	
	%\author{A. A. Morgan, V. Zhelyazkova, S. D. Hogan}
	\author{A. A. Morgan}
    \author{V. Zhelyazkova}
        \altaffiliation{Present address: Laboratorium f\"ur Physikalische Chemie, ETH Z\"urich, CH-8093 Z\"urich, Switzerland}
    \author{S. D. Hogan}
	%\email[]{alexandre.morgan.15@ucl.ac.uk}
	%\thanks{}
	\affiliation{Department of Physics and Astronomy, University College London, London, Gower Street, WC1E 6BT, United Kingdom}
	
	
	\date{\today}
	
	\begin{abstract}
		Circular Rydberg states with $n=70$ have been prepared in helium using a modified version of the crossed-fields method. This approach to the preparation of high-$n$ circular Rydberg states overcomes limitations of the standard crossed-fields method which arise at this, and higher, values of $n$. The experiments were performed with atoms traveling in pulsed supersonic beams that were initially laser photoexcited from the metastable 1s2s$\,^3$S$_1$ level to the 1s73s$\,^3$S$_1$ level by resonance-enhanced two-color two-photon excitation in a magnetic field of 16.154~G. These excited atoms were then polarized using a perpendicular electric field of 0.844~V/cm, and transferred by a pulse of microwave radiation to the state that, when adiabatically depolarized, evolves into the $n=70$ circular state in zero electric field. The excited atoms were detected by state-selective electric field ionization. Each step of the circular state preparation process was validated by comparison with the calculated atomic energy level structure in the perpendicular electric and magnetic fields used. Of the atoms initially excited to the 1s73s$\,^3$S$_1$ level, $\sim80$\%  were transferred to the $n=70$ circular state. At these high values of $n$, $\Delta n = 1$ circular-to-circular Rydberg state transitions occur at frequencies below 20~GHz. Consequently, atoms in these states, and the circular state preparation process presented here, are well suited to hybrid cavity QED experiments with Rydberg atoms and superconducting microwave circuits.
	\end{abstract}
	
	\pacs{}
	%\keywords{}
	
	\maketitle
	
	\section{\label{sec:introduction}Introduction}
	
	Circular Rydberg states are excited electronic states of atoms or molecules with a high principal quantum number, $n$, and maximal electron orbital angular momentum and azimuthal quantum numbers, $\ell=(n-1)$ and $|m_{\ell}|=(n-1)$, respectively~\cite{Hulet1983RydbergAtoms}. For values of $n$ larger than 50, $\Delta n=+1$ single-photon transitions between circular states occur at microwave frequencies below 50~GHz, and exhibit electric dipole transition moments in excess of $1800\,e a_0$. The selection rules for these transitions, involving the single excited Rydberg electron, require that $\Delta\ell=\pm1$ and $\Delta m_{\ell}=0,\pm1$, and therefore only allow spontaneous emission from one circular state to the next-highest-lying circular state, i.e., by a $\Delta n=-1$ transition. Because such spontaneous emission occurs in the microwave region of the electromagnetic spectrum, the corresponding Einstein A coefficients are typically less than 30~s$^{-1}$. Consequently the fluorescence lifetimes of circular Rydberg states with $n>50$ are greater than 30~ms. Pairs of circular states with values of $n$ that differ by 1 can therefore be considered quasi-two-level systems with strong electric dipole transitions at microwave frequencies. 
	
	Circular Rydberg states do not exhibit first-order Stark energy shifts, thus for each value of $n$, they represent the sublevels with the lowest sensitivity to electric fields~\cite{Gallagher1994RydbergAtoms}. The quadratic Stark energy shifts of these states are characterized by their static electric dipole polarizabilities, which for the circular state with $n=50$ [$n=70$] is $2.69\times10^{-31}$~C\,m$^2/$V$~\equiv~2.03$~MHz/(V/cm)$^2$ [$2.00\times10^{-30}$~C\,m$^2/ \mathrm{V} \equiv15.11$~MHz/(V/cm)$^2$]. These low static electric dipole polarizabilities contribute to long coherence times for superpositions of circular states which differ in $n$ by 1.
	
	The quasi-two-level character of $\Delta n=1$ transitions between circular Rydberg states has meant that they have long played an important role in microwave cavity quantum electrodynamics (cavity QED) experiments~\cite{Raimond2001, Walther2006CavityQED}. They were exploited, for example, in the realization of the single-atom maser~\cite{Meschede1985}, in early tests of the Jaynes-Cummings model of quantum optics~\cite{Rempe1987CollapseRevival}, and in quantum non-demolition (QND) detection of single microwave photons in three-dimensional superconducting microwave cavities~\cite{haroche2007BirthAndDeath}. 
	
	Recent developments in quantum enhanced electrometry~\cite{Haroche2017SensitiveElectrometer}, quantum simulation with arrays of cold Rydberg atoms~\cite{Haroche2018QuantumSimulation}, and hybrid quantum optics involving gas-phase Rydberg atoms and chip-based superconducting microwave circuits~\cite{Rabl2006, Hogan2012DrivingWaveguide, Pritchard2014HybridAtomPhoton, Rabl2017CherenkovPhotons} are expected to benefit from the use of circular Rydberg states. However, the requirements of these experiments place constraints on aspects of the circular state preparation process, and on the values of $n$, which were not encountered in previous microwave cavity QED experiments. For example, the use of circular Rydberg states for quantum enhanced electric field sensing~\cite{Haroche2017SensitiveElectrometer} requires fast and accurate state preparation. This has recently been addressed from the perspective of quantum optimal control theory for the case of the circular state with $n=51$ in rubidium~\cite{Patsch2018FastCircularization}. In the development of hybrid systems for quantum optics and quantum information processing, the optimal operating frequencies of two-dimensional superconducting microwave circuits and resonators lie below 20~GHz~\cite{Goppl2008CoplanarElectrodynamics}. To resonantly couple atoms in circular Rydberg states to these devices it is therefore desirable to prepare such states with values of $n\geq70$ for which $\Delta n=+1$ single-photon microwave transitions occur at below 20~GHz. In this setting, the large electric dipole transition moments of circular-to-circular state transitions at these high values of $n$ offer opportunities for strong coupling to single microwave photons in the resonators, while the low static electric dipole polarizability of the states minimizes their sensitivity to stray electric fields emanating from the chip surfaces. The desire to minimize time-dependent changes in these fields caused by adsorbates on cryogenic chip surfaces has motivated the use of helium atoms, rather than alkali metal atoms, in several of these experiments~\cite{Hogan2012DrivingWaveguide, Thiele2014ManipulatingRydberg}. The smaller ground state static electric dipole polarizability of helium, $\alpha_{\mathrm{He}}\sim1.38$~au~\cite{Mitroy2010TheoryApplicationsAtomic} ($1~\mathrm{au} = 1.648\,773\times10^{-41}$~C\,m$^2$/V), compared to those of the alkali metal atoms, e.g., $\alpha_{\mathrm{Rb}}\sim319$~au~\cite{Holmgren2010AbsoluteRatioPolarizability}, reduces adsorption, and results in smaller changes in the stray surface electric fields if adsorption does occur~\cite{hogan18a}. 
	
	The high orbital angular momentum of circular Rydberg states precludes direct laser photoexcitation from an atomic or molecular ground state. However, these states can be prepared by (1) adiabatic microwave transfer following laser photoexcitation of low-$\ell$ Rydberg states \cite{Hulet1983RydbergAtoms, Liang1986CircularStateSpectroscopy, Cheng1994ProductionCircularStates}, or (2) laser photoexcitation in the presence of perpendicularly crossed electric and magnetic fields, followed by adiabatic extraction from these fields \cite{DelandeGay1988CircularStates, HareGrossGoy1988CrossedFields}. Both of these approaches are suited to the preparation of circular states with values of $n$ up to $\sim60$. However, because of the increased sensitivity of the preparation process to uncanceled stray electric fields and to electric field noise, they are more challenging to implement at higher values of $n$. 
	
	The standard crossed-fields method for preparing circular Rydberg states offers greater flexibility in the range of values of $n$ that can be addressed in one apparatus than the adiabatic microwave transfer method. It has been demonstrated in experiments with hydrogen~\cite{Lutwak1997CircularHydrogen}, helium~\cite{Zhelyazkova2016PreparationCircular}, lithium~\cite{HareGrossGoy1988CrossedFields}, rubidium~\cite{Brecha1993CircularRydbergStates, Anderson2013ProductionColdRydberg} and barium~\cite{Cheret1989BariumCircular}, and forms the basis of the approach to the preparation of circular states with values of $n\geq70$ presented here. The modified crossed-fields method presented can be implemented with higher efficiency and state selectivity at these high values of $n$ than the standard approach, and may be applied to the preparation of circular Rydberg states with values of $n$ in excess of 100 in atoms or molecules. 
	
	The high-$n$ circular Rydberg states in helium prepared in the experiments described here, are ideally suited for use in hybrid cavity QED when they are coupled to chip-based superconducting microwave resonators, with applications as long-coherence-time quantum memories~\cite{Rabl2006}, and in studies of strong coupling to slow-light Cherenkov photons~\cite{Rabl2017CherenkovPhotons}. They may also offer opportunities for sensitive quantum non-demolition detection \cite{Braginksy1980, Unruh1978AnalysisQND} of cold polar ground state molecules by Ramsey interferometry of dipole-dipole interactions in the dispersive regime. This would be of particular relevance in the detection of single alkali-metal dimers prepared in optical tweezers \cite{Liu2018BuildingOneMolecule} for which pure rotational transitions between the lowest-lying rotational states occur at frequencies below $\sim$~25~GHz \cite{Aymar2005CalculationPermanentDipole}. Such methods of quantum non-demolition detection go beyond other recently proposed schemes in which Rydberg atoms are exploited for the detection~\cite{Zeppenfeld2017NondestructivePolar} and interrogation~\cite{Kuznetsova2016RydbergNondestructiveReadout} of cold polar molecules. 
	
	In the following, the apparatus in which the experiments were carried out is first described in Section~II, and the theoretical background to the calculations performed to optimize the circular state preparation process and aid in the analysis of the experimental data, is outlined in Section~III. In Section~IV, the modified crossed-field method for the preparation of high-$n$ circular Rydberg states is presented, before the experimental implementation of this approach to the preparation of the $n=70$ circular Rydberg state in helium is reported in Section~V. In Section~VI and VII, the requirements for adiabatic evolution of the excited states during the preparation process, and considerations relating to the appropriate magnetic field strengths for the implementation of this approach to circular state preparation at other values of $n$ and in other atoms and molecules are discussed. In Section~VIII the application of the high-$n$ circular Rydberg states, prepared in the experiments described here, in hybrid cavity QED is discussed before conclusions are drawn in Section~IX. 
	
	% ------------------------------------------------
	%               Experimental setup
	% ------------------------------------------------
	
	\section{\label{sec:experiment}Experiment}
	
\begin{figure}
		\includegraphics[width=0.5\textwidth]{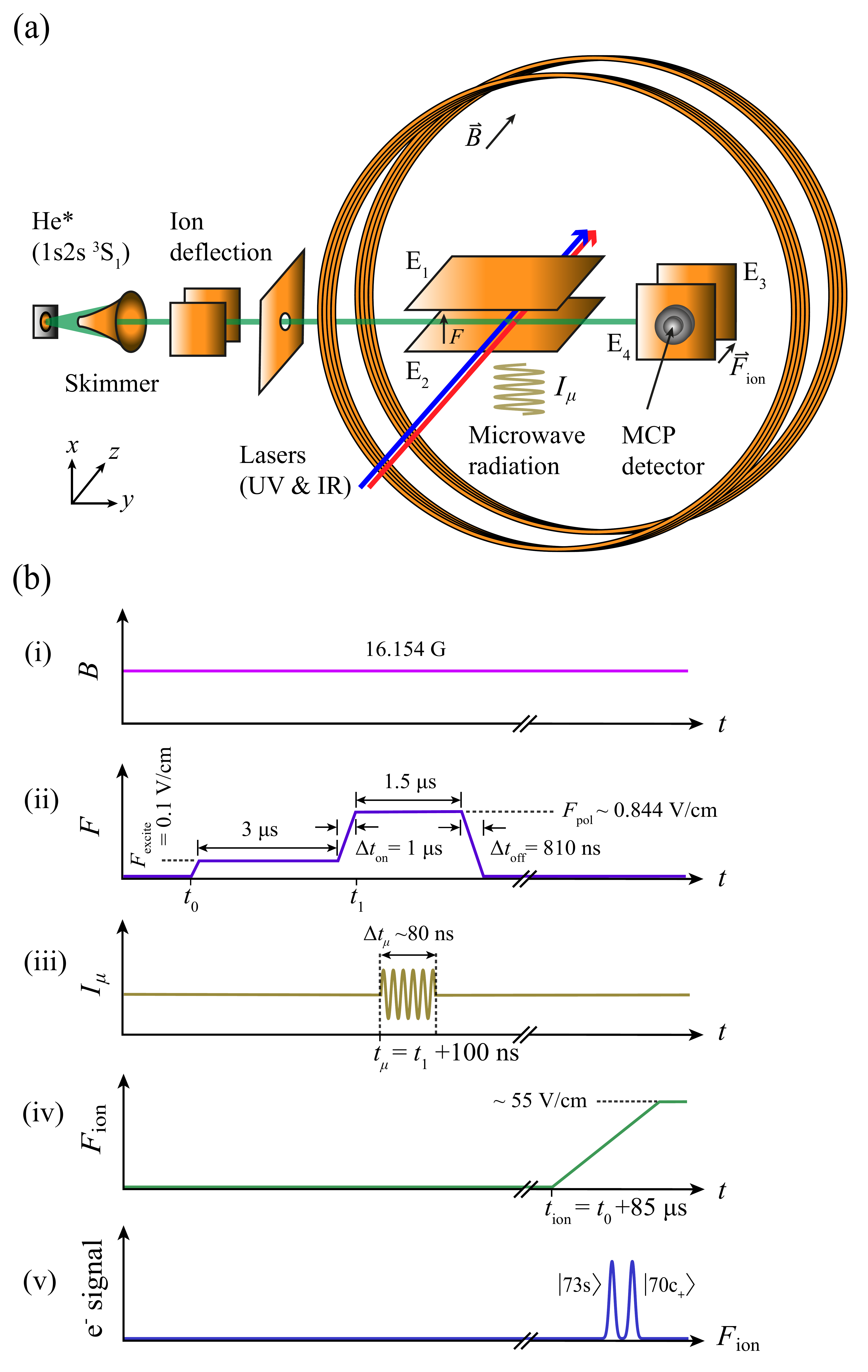}
		\caption{\label{fig:experimental-setup} Schematic diagrams of (a) the experimental setup, and (b) the temporal sequence of each cycle of the experiment: (i) background magnetic field, $B = |\vec{B}| = B_z$, (ii) electric field in the photoexcitation region, $F = |\vec{F}| = F_x$, (iii) pulse of microwave radiation with intensity $I_{\mathrm{\mu}}$, (iv) time-dependent ionization electric field, and (v) electron signal recorded at the microchannel plate (MCP) detector. }
	\end{figure}

	A schematic diagram of the experimental apparatus, similar to that described in Ref.~\cite{Zhelyazkova2016PreparationCircular}, is shown in Figure~\ref{fig:experimental-setup}(a). A pulsed supersonic beam of helium in the metastable 1s2s$\,^3$S$_1$ level and traveling with a mean longitudinal speed of $\sim 2000$~m/s, was generated in a DC electric discharge at the exit of a pulsed valve operated at a repetition rate of 50~Hz \cite{Halfmann2000SourcePulsedBeam}. After passing through a skimmer with a 2~mm diameter aperture, ions generated in the discharge were removed from the beam by electrostatic deflection. The remaining neutral atoms then entered the region between the two parallel metal plates labeled E$_1$ and E$_2$ in Figure~\ref{fig:experimental-setup}(a). These plates had dimensions of 70$\times$100 mm in the $zy$ plane and were separated in the $x$ dimension by 13.0~mm. In this part of the apparatus a homogeneous magnetic field of $|\vec{B}| = 16.154$~G acting in the $z$ direction was generated by a pair of 160~mm diameter coils operated in a Helmholtz configuration and located outside the vacuum chamber. Between plates E$_1$ and E$_2$, the beam of atoms was intersected by co-propagating ultraviolet (UV) and infrared (IR) laser beams. The UV laser was frequency stabilized to drive the 1s2s$\,^3$S$_1$~$\rightarrow$~1s3p$\,^3$P$_2$ ($\Delta M_{J} = +1$) transition at a wavelength of $\lambda_{\mathrm{UV}} = $~388.975~nm. The IR laser was set to lie resonant with 1s3p$\,^3$P$_2$~$\rightarrow$~1s$n$s$\,^3$S$_1$ transitions, which for $n=73$ occurs at a wavelength of $\lambda_{\mathrm{IR}} =$~785.833~nm. In the crossed-fields configuration used here the motional Stark effect, here an effective field of 32.3~mV/cm for a beam of atoms traveling at $\sim2000$~m/s perpendicular to a 16.154~G magnetic field, can be canceled by applying an equal but opposite offset field between E$_1$ and E$_2$~\cite{Elliott1995CancellationMotionalStark}. After laser photoexcitation, a sequence of electric field, $F$, and microwave, $I_{\mu}$, pulses was applied to transfer the atoms to the circular Rydberg states, as indicated in Figure~\ref{fig:experimental-setup}(b). Upon completion of the circular state preparation process the atoms traveled 170~mm down stream to the detection region of the apparatus between plates E$_3$ and E$_4$. At this point the atoms were detected by state-selective ionization using time-dependent electric fields, $\vec{F}_{\mathrm{ion}}$, orientated parallel to the background magnetic field. The ionized electrons were then accelerated through a hole in the electrode labeled E$_4$ and collected on a microchannel plate (MCP) detector.
	
	% ------------------------------------------------
	%    Stark map & modified crossed-fields method
	% ------------------------------------------------
	
	\section{\label{sec:theoretical-background}Theoretical background and numerical calculations}
	
	In the presence of perpendicularly crossed electric, $\vec{F} = (F_x,0,0)$, and magnetic, $\vec{B} = (0,0,B_z)$, fields, the Hamiltonian associated with a single Rydberg electron, excluding spin, has the form
	\begin{eqnarray}
	\label{eq:hamiltonian-zeeman-stark_general}
	\hat{H} & = \hat{H}_0 + \frac{\mu_B}{\hbar} \hat{\vec{B}} \cdot \hat{\vec{L}} + e \hat{\vec{F}} \cdot \hat{\vec{r}} \\
	\label{eq:hamiltonian-zeeman-stark}
	& = \hat{H}_0 + \frac{\mu_B}{\hbar} B_z \hat{L}_z + e F_x \hat{x},
	\end{eqnarray}
	where, in matrix form in the $| n, \ell, m_{\ell} \rangle$ basis, $\hat{H}_0$ is diagonal and represents the field-free level energies determined using the Rydberg formula with the appropriate quantum defects for the low-$\ell$ states, $\hat{\vec{L}}$ and $\hat{\vec{r}}$ are the total electron orbital angular momentum and radial position operators, respectively, $\mu_0$ is the Bohr magneton, $\hbar$ is the reduced Planck constant, and $e$ is the electron charge. The $n$-dependent quantum defects of the triplet levels in helium with $\ell \leq 5$, as determined from Ref.~\cite{Drake1999HighPrecision}, were employed in all calculations reported here, and are listed in Table~\ref{tab:quantum-defects} for $n=$~73. When converted to spherical polar coordinates, the electric field contribution to the Hamiltonian in Equation~(\ref{eq:hamiltonian-zeeman-stark}) can be separated into angular and radial components. For the case of interest here the angular integrals, which arise from the coupling of sublevels for which $\Delta \ell = \pm 1$ and $\Delta m_{\ell} = \pm 1$, can be expressed analytically following standard angular momentum algebra~\cite{Zhelyazkova2016PreparationCircular}. However, the corresponding radial integrals must be computed numerically. This was done using the Numerov method~\cite{Gallagher1994RydbergAtoms, Zimmerman1979StarkStructure}. For the calculations at values of $n$ close to 70, Numerov integration step sizes of $h = 0.005 \, a_0$ were required to achieve convergence of the eigenvalues of the resulting matrices to $\sim$1~MHz. These eigenvalues were determined by diagonalizing the complete Hamiltonian matrix corresponding to Equation~(\ref{eq:hamiltonian-zeeman-stark}).
	
	The significance of the diamagnetic interaction~\cite{Kash1977Diamagnetic} was tested in the calculations and found to result in frequency shifts of less than 1~MHz. Since these shifts are smaller than the spectral widths in the experimental data, this second order correction was not included in the calculated data presented here. For the high values of $n$ of interest here, the largest zero-field fine-structure splittings, i.e., those of the 1s$n$p$\,^3$P$_J$ levels, are $\sim \,$0.5~MHz~\cite{Drake1999HighPrecision,deller18a}. Therefore, in the calculations used to interpret and assign the measured microwave spectra, spin-orbit contributions were neglected. However, their role in the excited state dynamics in the time-dependent electric fields used in the experiments is discussed in Section~VI.
	
	\begin{table}[b]%The best place to locate the table environment is directly after its first reference in text
		\caption{\label{tab:quantum-defects} Quantum defects of $n=73$ triplet Rydberg states in helium.}
		\begin{ruledtabular}
			\begin{tabular}{c}
				\textrm{Quantum defects, $\delta_{n, \ell}$} \\
				\colrule                                     
				$\delta_{73,0} = 0.296\,664$                 \\
				$\delta_{73,1} = 0.068\,357$                 \\
				$\delta_{73,2} = 0.002\,890$                 \\
				$\delta_{73,3} = 0.000\,447$                 \\
				$\delta_{73,4} = 0.000\,127$                 \\
				$\delta_{73,5} = 0.000\,049$                 \\
			\end{tabular}
		\end{ruledtabular}
	\end{table}
	
	In the following we denote the $m_{\ell} = \pm(n-1)$ circular state at each value of $n$ as $| n\mathrm{c_{\pm}} \rangle$. The $\ell$-mixed Stark states are distinguished from the pure-$\ell$ states into which they adiabatically evolve in zero electric field with a prime, e.g., $| n\mathrm{c'_+} \rangle$ for the outer-most low-field-seeking Stark state in a non-zero electric field. 
	
\begin{figure}[h!]
		\includegraphics[width=0.5\textwidth]{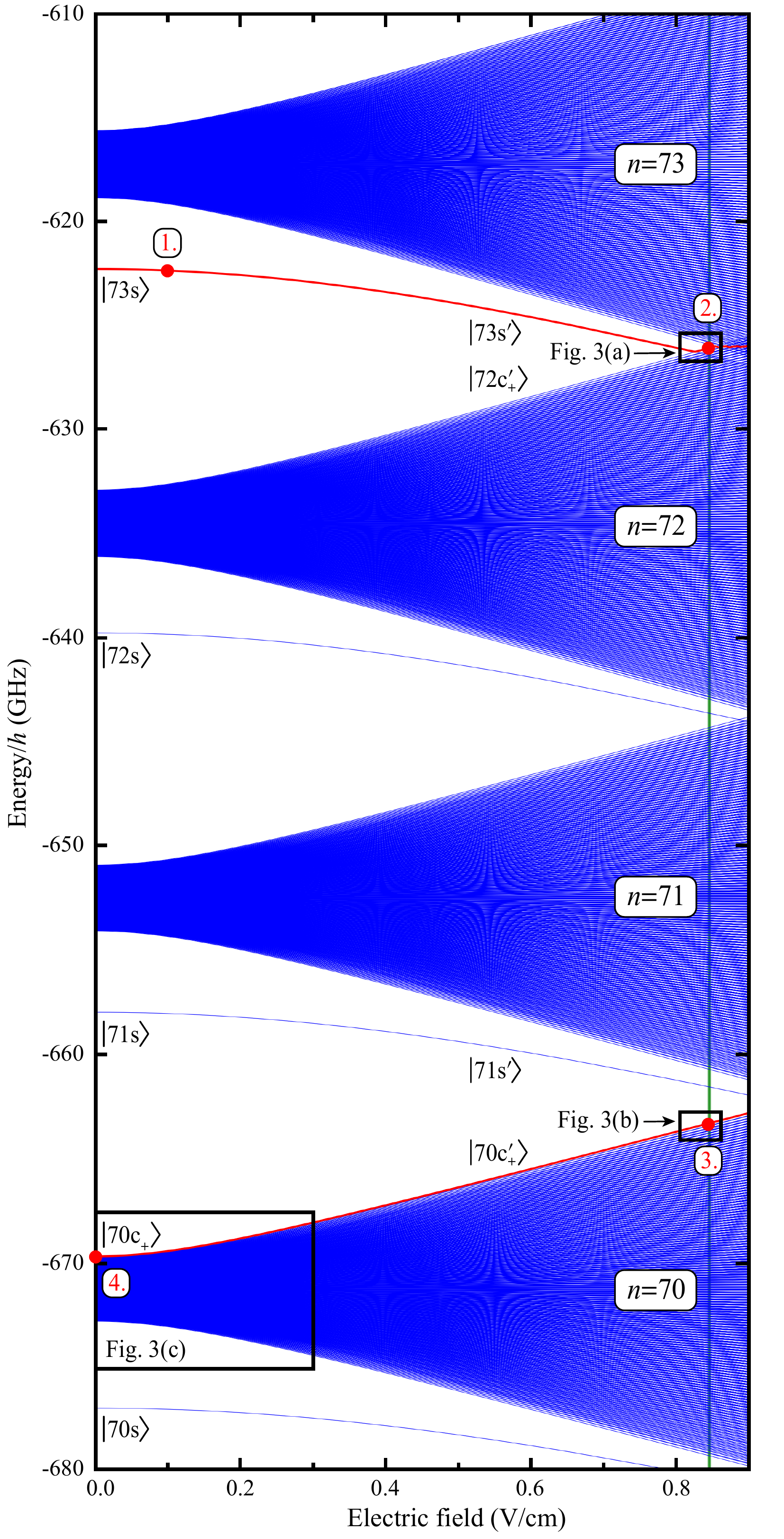}
		\caption{\label{fig:stark-map-full} Energy level diagram of the triplet Rydberg states in helium with $70 \leq n \leq 73$ in a magnetic field $B_z = 16.154$~G and perpendicular electric fields. The zero electric field $| n=73, \ell=0, m_{\ell}=0 \rangle \equiv | 73\mathrm{s} \rangle$ and $| n=70, \ell=69, m_{\ell}=+69 \rangle \equiv | 70\mathrm{c_+} \rangle$ sublevels are highlighted in red. In the presence of the electric field these states are denoted $| 73\mathrm{s'} \rangle$ and $| 70\mathrm{c'_+}\rangle$, respectively (see text for details). The red points, labeled $1-4$, correspond to the steps of the circular state preparation process.}
	\end{figure}

	A calculated energy level diagram of the triplet $70 \leq n \leq 73$ Rydberg states of helium in a magnetic field of 16.154~G and perpendicular electric fields of up to 0.9~V/cm is shown in Figure~\ref{fig:stark-map-full}. For electric fields below the Inglis-Teller limit, where manifolds which differ in $n$ by 1 first cross, convergence to $\sim$~$\,$1~MHz in the calculated eigenvalues was achieved for bases containing all $| n, \ell, m_{\ell} \rangle$ sublevels within a frequency range of $\pm 50$~GHz around the energy of interest. Consequently, to accurately calculate the energy level structure at the avoided crossing where the $| 73\mathrm{s'} \rangle$ state joins the $n=$~72 manifold of $\ell$-mixed Stark states, i.e., the $| 72\mathrm{c'_+} \rangle$ state, a basis of states with $70 \leq n \leq 75$ was used. Likewise, to accurately calculate the energy level structure surrounding the $| 70\mathrm{c'_+} \rangle$ state in this same range of electric fields a basis with $68 \leq n \leq 73$ was used. This resulted in Hamiltonian matrices of dimensions $31555 \times 31555$ and $29839 \times 29839$, respectively.
	
	% ------------------------------------------------
	%       The modified crossed-fields method
	% ------------------------------------------------
	
	\section{\label{sec:circular-state-preparation}Circular state preparation process}
	
	In previously demonstrated implementations of the crossed-fields method for the preparation of circular Rydberg states \cite{DelandeGay1988CircularStates, HareGrossGoy1988CrossedFields, Zhelyazkova2016PreparationCircular}, the outermost low-field-seeking Stark state with $m_{\ell} = 0$ with respect to the electric field quantization axis, i.e., the $| n\mathrm{c'_+} \rangle$ state, was directly laser photoexcited. This was carried out at, or close to, the Inglis-Teller limit, where in non-hydrogenic systems this state generally acquires significant low-$\ell$ character. After photoexcitation, the electric field was adiabatically reduced to zero causing the excited state to evolve into the $| n\mathrm{c_+} \rangle$ circular state with maximal $m_{\ell}$ with respect to the magnetic field quantization axis.
	
	Although this original version of the crossed-fields method is quite general, it presents a significant challenge when used to prepare circular Rydberg states with very high values of $n$, i.e., $n\gtrsim~60$. The avoided crossings at the Inglis-Teller limit are sufficiently large at low values of $n$ that the individual Stark sublevels can be readily resolved in the photoexcitation process. However, for high values of $n$ these crossings reduce in size and the relevant levels, which possess large static electric dipole moments on either side of the avoided crossings, become increasingly sensitive to electric field noise \cite{HoganZhelyazkova2015OscillatingFields}. This makes it difficult to ensure complete state-selective preparation of individual high-$n$ sublevels by laser photoexcitation. Here, we present a modified version of the crossed-fields method which overcomes this problem. This is achieved by first performing the initial laser photoexcitation to an isolated low-$\ell$ (in this case $| n\mathrm{s} \rangle$) level in zero, or a weak electric field, where it can be efficiently excited and resolved. Then, after polarization of the atoms, by subsequently driving a microwave transition to transfer population between Stark sublevels with similar static electric dipole moments, the sensitivity of the circular state preparation process to electric field noise is reduced by more than an order of magnitude.
	
	\begin{figure}
		\includegraphics[width=0.5\textwidth]{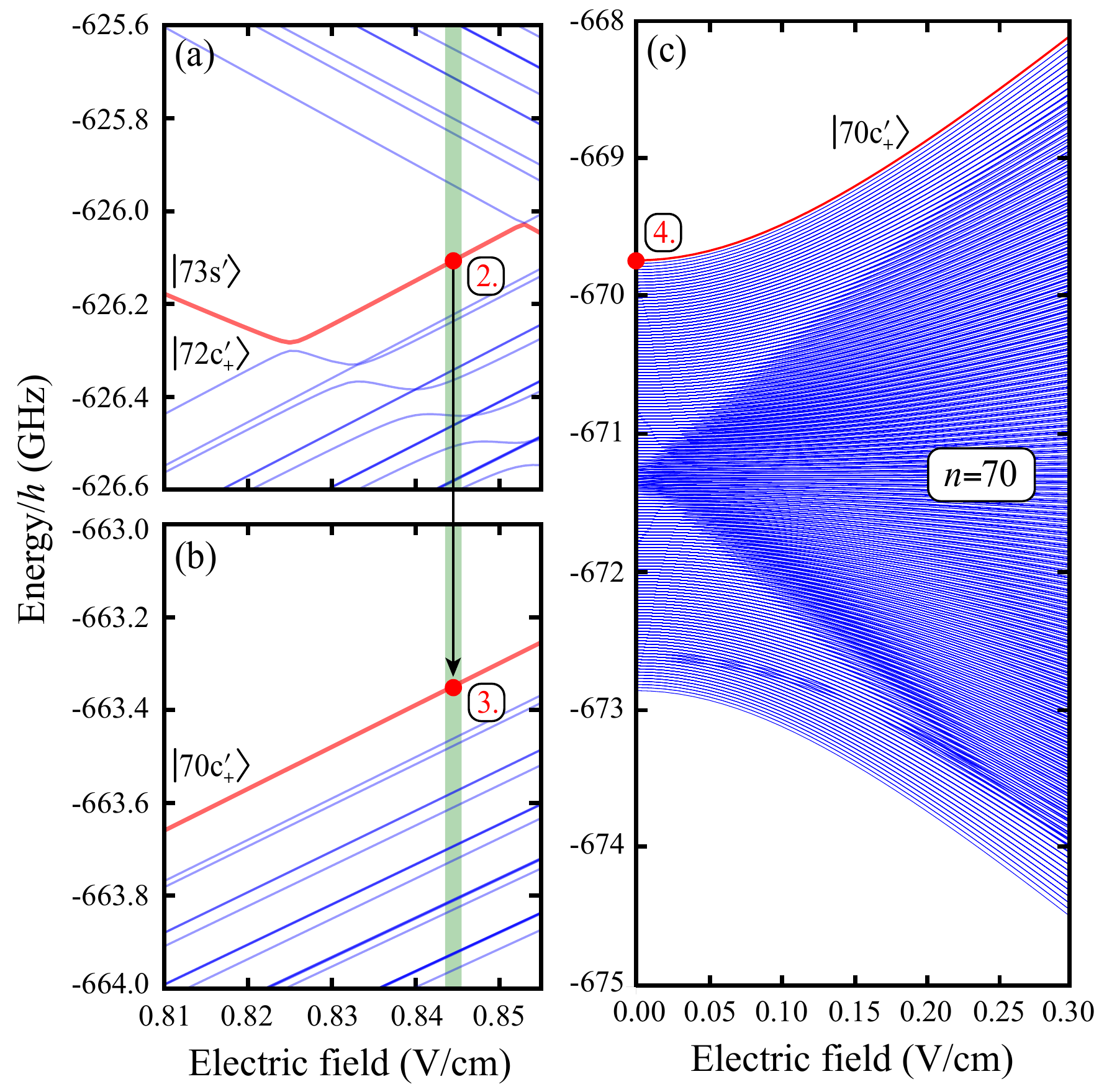}
		\caption{\label{fig:stark-map-zoom} Expanded views of the highlighted regions shown in Figure~\ref{fig:stark-map-full}. The arrow joining panels (a) and (b) indicates the $| 73\mathrm{s'} \rangle \rightarrow | 70\mathrm{c'_+}\rangle$ microwave transition at $\nu \approx 37.2312$~GHz (see text for details).}
	\end{figure}
	
	This modified crossed-fields method is implemented in 4 steps. In the case of the preparation of the $| 70\mathrm{c_+}\rangle$ state in helium presented here, this involves the following:
	
	\begin{itemize}
		\item[1.] Laser photoexcitation in a weak electric field of $F_{\mathrm{excite}} = 0.1$~V/cm from the  1s2s$\,^3$S$_1$ level to the 1s73s$\,^3$S$_1$ level using a resonance-enhanced two-color two-photon scheme through the  1s3p$\,^3$P$_2$ intermediate level (red point 1 in Figure~\ref{fig:stark-map-full}).
		
		\item[2.] Polarization of the excited atoms by adiabatically increasing the electric field strength in a time $\Delta t_{\mathrm{on}} = 1 \, \mu$s to $F_{\mathrm{pol}} = 0.844$~V/cm [red point 2 in Figure~\ref{fig:stark-map-full} and \ref{fig:stark-map-zoom}(a)]. This field is beyond the first avoided crossing between the $| 73\mathrm{s'} \rangle$ and $| 72\mathrm{c'_+}\rangle$ states, and results in the static electric dipole moment of the polarized $| 73\mathrm{s'} \rangle$ state being orientated anti-parallel to the electric field vector.
		
		\item[3.] At point 2 in the Stark map the single-photon transition from the $| 73\mathrm{s'} \rangle$ state to the $| 70\mathrm{c'_+}\rangle$ state has a significant electric dipole transition moment because of the large overlap between the wavefunctions of the two states which posses similar static electric dipole moments [compare the energy level structure at red point 2 in Figure~\ref{fig:stark-map-zoom}(a) with that at red point 3 in Figure~\ref{fig:stark-map-zoom}(b)]. Therefore a short, $\Delta t_{\mu} \sim 80$~ns pulse of microwave radiation, resonant with the $| 73\mathrm{s'} \rangle \rightarrow | 70\mathrm{c'_+}\rangle$ transition at $\nu = 37.2312$~GHz when $F_\mathrm{pol} = 0.844$~V/cm, is applied to transfer population to the $| 70\mathrm{c'_+}\rangle$ state. In this particular case, the difference between the static electric dipole moments of the $| 73\mathrm{s'} \rangle$ and $| 70\mathrm{c'_+}\rangle$ states is $\Delta \mu = \mu_{70\mathrm{c'_+}} - \mu_{73\mathrm{s'}} = 1072$ D [$\Delta \mu/h = -540$~MHz/(V/cm)] and is almost constant over an electric field range of $\sim 0.02$~V/cm, between the first and second avoided crossings encountered by the $| 73\mathrm{s'} \rangle$ state. Therefore the transition has low sensitivity to DC electric field inhomogeneities and low-frequency electric field noise.
		
		\item[4.] After preparation of the $| 70\mathrm{c'_+}\rangle$ state the electric field is switched off in a time $\Delta t_{\mathrm{off}} = 810$~ns, corresponding to a rate of $\mathrm{d}F/\mathrm{d}t~=~1.03$~(V/cm)/$\mu$s, to allow the atoms to adiabatically evolve into the pure $| 70\mathrm{c_+}\rangle$ circular state in the magnetic field.
	
	\end{itemize}
	
	% ------------------------------------------------
	%                     Results
	% ------------------------------------------------
	
	\section{\label{sec:results}Experimental implementation}
	
	% ------------------------------------------------
	%        Ramped electric field ionization
	% ------------------------------------------------
	
\begin{figure}
		\includegraphics[width=0.5\textwidth]{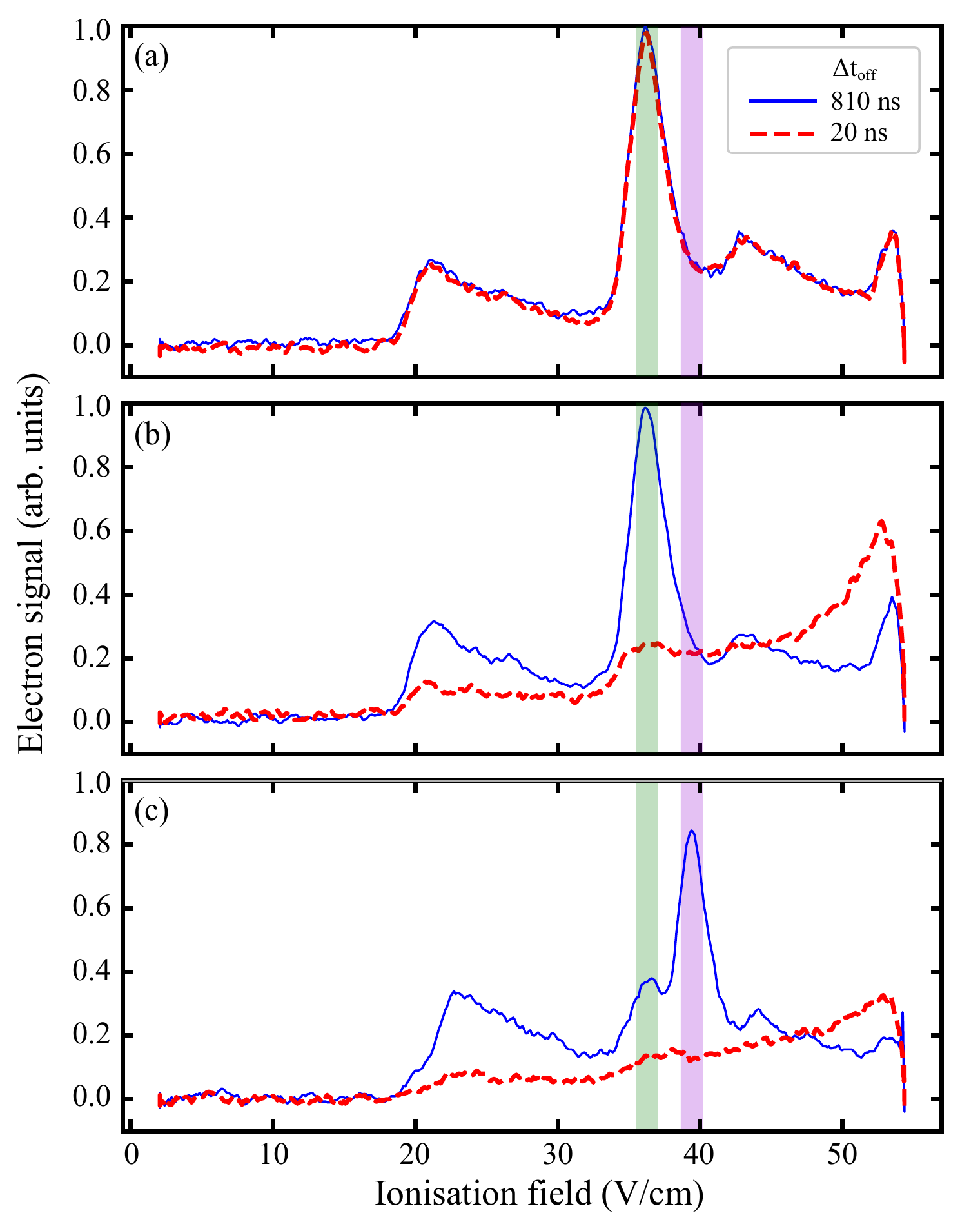}
		\caption{\label{fig:mcp-adiabatic} Electron signals recorded upon electric field ionization of the Rydberg atoms in parallel electric and magnetic fields. (a) The polarization electric field was adiabatically increased in a time $\Delta t_{\mathrm{on}} = 1 \mu$s to $F_{\mathrm{pol}}=0.80$~V/cm, before the first avoided crossing of the $| 73\mathrm{s'} \rangle$ state, and subsequently switched off. (b) The polarization electric field was adiabatically increased to $F_{\mathrm{pol}}=0.844$~V/cm, after the first avoided crossing, and subsequently switched off. (c) The polarization electric field was adiabatically increased to $F_{\mathrm{pol}}=0.844$~V/cm, then a microwave $\pi$-pulse, resonant with the $| 73\mathrm{s'} \rangle \rightarrow | 70\mathrm{c'_+}\rangle$ transition ($\nu \approx 37.2312$~GHz), was applied, before $F$ was switched off. In each case, the electric field was switched off in a time $\Delta t_{\mathrm{off}} = 810$ ns (continuous blue curve), and $\Delta t_{\mathrm{off}} = 20$ ns (dashed red curve). The green (left) and purple (right) shaded regions in each panel are centered on the electric fields at which the $| 73\mathrm{s} \rangle$ and $| 70\mathrm{c_+}\rangle$ states ionize, respectively.}
	\end{figure}

	To implement this modified crossed-fields method for the preparation of the $| 70\mathrm{c_+}\rangle$ circular state, it was first necessary to ensure that the avoided crossing between the $| 73\mathrm{s'} \rangle$ and $| 72\mathrm{c'_+}\rangle$ states, at a field of $F_{\mathrm{ac}} \simeq 0.825$~V/cm, shown in Figure~\ref{fig:stark-map-zoom}(a), was traversed adiabatically. This was checked experimentally by first increasing the electric field from $F_{\mathrm{excite}} = 0.1$~V/cm to $F_{\mathrm{pol}}=0.8$~V/cm (just below $F_{\mathrm{ac}}$) in a time $\Delta t_{\mathrm{on}} = 1~\mu$s. This electric field was then switched off in a time $\Delta t_{\mathrm{off}} = 810$~ns before the excited atoms were subsequently detected by state-selective ionization in parallel electric and magnetic fields. The resulting electron signal is shown as the continuous blue curve in Figure~\ref{fig:mcp-adiabatic}(a). The first feature that appears at an ionization field of $\sim~21.4$~V/cm in this data set, corresponds to the adiabatic electric field ionization of the $| 73\mathrm{s} \rangle$ state. The other features, including the dominant one at $\sim~36.2$~V/cm, represent distinct non-adiabatic ionization pathways of this same state. The adiabaticity of the ionisation dynamics in the nominally parallel electric and magnetic fields in the detection region of the apparatus are particularly sensitive to the alignment of these fields~\cite{Brecha1993CircularRydbergStates}.  In recording the data presented as the continuous blue curve in Figure~\ref{fig:mcp-adiabatic}(a) the atoms were not subjected to electric fields beyond the Inglis-Teller limit before they underwent electric field ionization. Therefore, reducing the switch-off time to $\Delta t_{\mathrm{off}} = 20$~ns [dashed red curve in Figure~\ref{fig:mcp-adiabatic}(a)] did not affect the field ionization signal.
	
	This experiment was repeated with $F_{\mathrm{pol}}$ increased to $0.844$~V/cm, between the first and second avoided crossings encountered by the $| 73\mathrm{s'} \rangle$ state. In this case a distinct difference was seen in a similar pair of measurements. When the field was switched off in $\Delta t_{\mathrm{off}} = 810$~ns [continuous blue curve in Figure~\ref{fig:mcp-adiabatic}(b)], the electron signal remained the same as that in Figure~\ref{fig:mcp-adiabatic}(a). However, when the field was switched off in $\Delta t_{\mathrm{off}} = 20$~ns, significant differences were observed indicating the population of a broader range of hydrogenic Stark states [dashed red curve in Figure~\ref{fig:mcp-adiabatic}(b)]. This demonstrates that switching the field from $F_{\mathrm{excite}} = 0.1$~V/cm to $F_{\mathrm{pol}}=0.844$~V/cm in a time $\Delta t_{\mathrm{on}} = 1~\mu$s, and subsequently switching it off in a time $\Delta t_{\mathrm{off}} = 810$~ns ensures adiabatic traversal of the first avoided crossing between the $| 73\mathrm{s'} \rangle$ and $| 72\mathrm{c'_+}\rangle$ states, but that a switch-off time of, e.g., $\Delta t_{\mathrm{off}} = 20$~ns results in diabatic traversal.
	
\begin{figure}
		\includegraphics[width=0.5\textwidth]{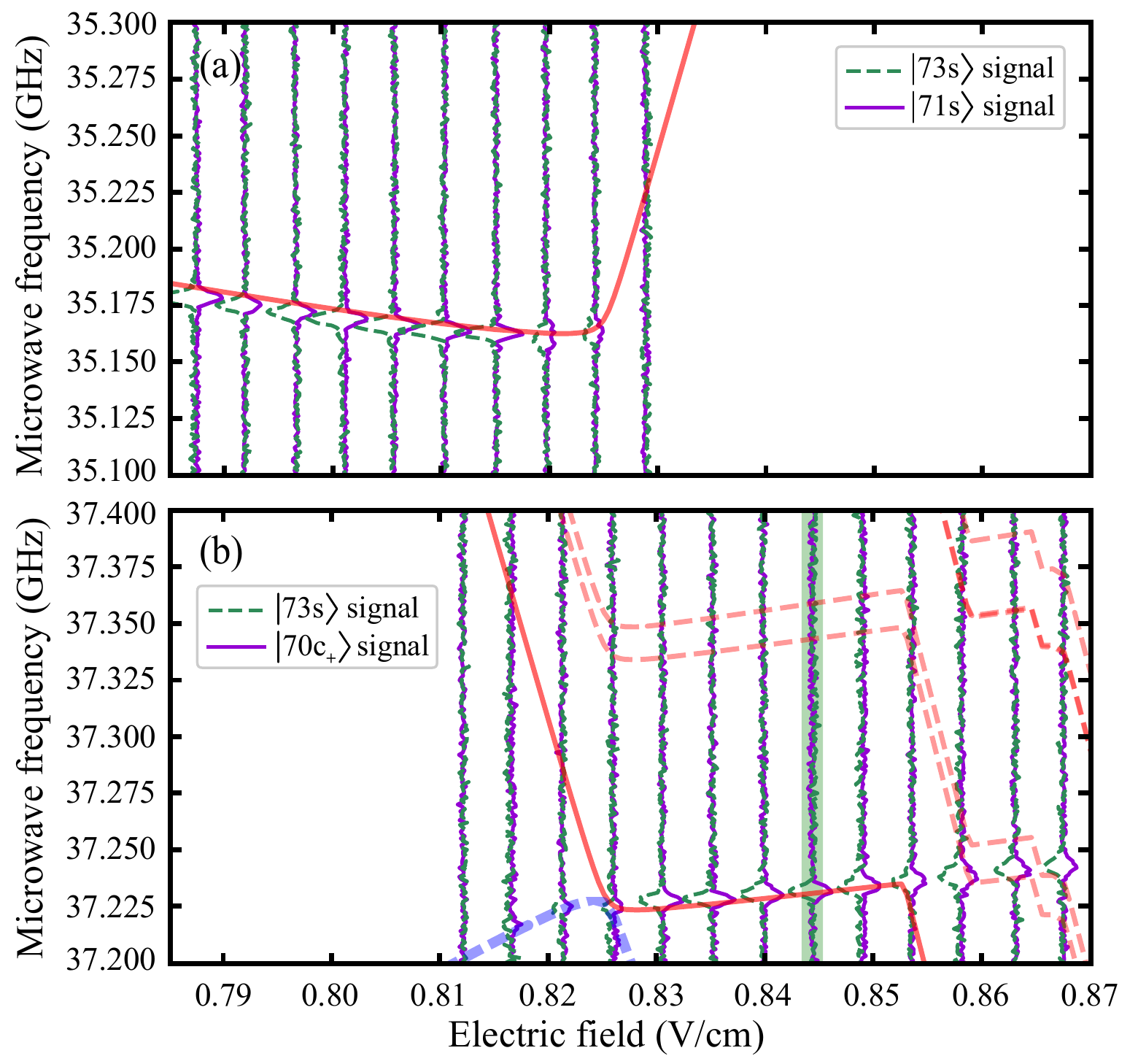}
		\caption{\label{fig:transition-spectra} Microwave spectra showing (a) the depletion and increase of the integrated electron signal in the $| 73\mathrm{s} \rangle$ (dashed curve) and $| 71\mathrm{s} \rangle$ (continuous curve) ionization windows, respectively. The continuous red curve represents the calculated $| 73\mathrm{s'} \rangle \rightarrow | 71\mathrm{s'} \rangle$ transition frequencies. (b) Depletion and increase of the integrated electron signal in the $| 73\mathrm{s} \rangle$ (dashed curve) and $| 70\mathrm{c_+}\rangle$ (continuous curve) ionization windows, respectively. Calculated frequencies for the $| 73\mathrm{s'} \rangle \rightarrow | 70\mathrm{c'_+}\rangle$ transition (continuous red curve), along with frequency intervals to other lower (dashed red curves) and higher (dashed blue curve) energy states, are shown.}
	\end{figure}

	Having identified the necessary conditions for adiabatic polarization of the $| 73\mathrm{s} \rangle$ state such that it possesses a static electric dipole moment of similar magnitude and orientation to the target $| 70\mathrm{c'_+}\rangle$ state in the same electric field, the process of resonant single-photon microwave transfer to this target state was characterized for different values of $F_{\mathrm{pol}}$. First, depletion spectra of the $| 73\mathrm{s'} \rangle$ state were recorded by integrating the signal in the window indicated by the green shaded region at $\sim~36.2$~V/cm in Figure~\ref{fig:mcp-adiabatic}. By driving the $| 73\mathrm{s'} \rangle \rightarrow | 71\mathrm{s'} \rangle$ transition the field at which the first avoided crossing between the $| 73\mathrm{s'} \rangle$ and $| 72\mathrm{c'_+}\rangle$ states occurs was determined. This is signified by the reduction of the transition strength to zero. Having identified the ionization field of the $| 71\mathrm{s'} \rangle$ state, spectra corresponding to the appearance of the integrated $| 71\mathrm{s'} \rangle$ signal were also recorded. These two sets of spectra are shown in Figure~\ref{fig:transition-spectra}(a) as the dashed green and continuous purple curves, respectively. The calculated $| 73\mathrm{s'} \rangle \rightarrow | 71\mathrm{s'} \rangle$ transition frequencies, obtained from the data in Figure~\ref{fig:stark-map-zoom}, are shown as the continuous red curve and are in good quantitative agreement with the experimental data to within $\sim~10$~MHz.
	
	In the experimental data presented in Figure~\ref{fig:transition-spectra}(a), the spectral intensity of the $| 73\mathrm{s'} \rangle \rightarrow | 71\mathrm{s'} \rangle$ transition tends towards zero immediately after the avoided crossing between the $| 73\mathrm{s'} \rangle$ and $| 72\mathrm{c'_+}\rangle$ states. This is because the sign of the static electric dipole moment of the $| 73\mathrm{s'} \rangle$ state changes at the avoided crossing, reducing the overlap of the electronic wavefunction of this state with that of the $| 71\mathrm{s'} \rangle$ state. On the other hand, this change in sign of the dipole moment increases the overlap with the wavefunction of the $| 70\mathrm{c'_+}\rangle$ state. The change in the spectral intensity of the $| 73\mathrm{s'} \rangle \rightarrow | 70\mathrm{c'_+}\rangle$ transition around the avoided crossing is seen in Figure~\ref{fig:transition-spectra}(b). The integration window used to monitor the signal from the $| 70\mathrm{c'_+}\rangle$ state in these spectra is indicated by the shaded purple region at $\sim~39.4$~V/cm in Figure~\ref{fig:mcp-adiabatic}. In parallel electric and magnetic fields, as used in the field ionization step of the experiments presented here, the electric field ionization rate of the  $| 70\mathrm{c'_+}\rangle$ state can be determined from the hydrogenic ionization rates corrected for the reduced mass of helium \cite{Damburg1979HydrogenAtom, Damburg1983RydbergStates}. For an ionization rate of $10^8$~s$^{-1}$ ($10^9$~s$^{-1}$), the ionization electric field of this state is expected to be 37.8~V/cm (38.3~V/cm).
	
	The calculated $| 73\mathrm{s'} \rangle \rightarrow | 70\mathrm{c'_+}\rangle$ transition frequencies are shown in Figure~\ref{fig:transition-spectra}(b) as the continuous red curve and agree with the experimental data to within $\sim~10$~MHz. The dashed red and blue curves indicate frequency intervals to other states below and above the $| 73\mathrm{s'} \rangle$ state, respectively. From the spectra in Figure~\ref{fig:transition-spectra} the avoided crossing between the $| 73\mathrm{s'} \rangle$ and $| 72\mathrm{c'_+}\rangle$ states was identified to occur at a field of 0.825~V/cm. In fields between 0.825~V/cm and 0.853~V/cm the similarity of the static electric dipole moments of the $| 73\mathrm{s'} \rangle$ and $| 70\mathrm{c'_+}\rangle$ states, $-18984$~D [9.56~GHz/(V/cm)] and $-17912$~D [9.02~GHz/(V/cm)], respectively, gives rise to the comparatively small change in the $| 73\mathrm{s'} \rangle \rightarrow | 70\mathrm{c'_+}\rangle$ transition frequency over this range of fields. An electric field of 0.844~V/cm was chosen for the implementation of the microwave transfer step (step~3) of the circular state preparation process. This was performd using a pulse of microwave radiation at a frequency of $\nu = 37.2312$~GHz [green shaded vertical band in Figure~\ref{fig:transition-spectra}(b)]. The effect of an 80~ns $\pi$-pulse followed by a depolarization time $\Delta t_{\mathrm{off}} = 810$~ns in which the excited atoms adiabatically evolved into the $| 70\mathrm{c_+}\rangle$ state, can be seen in the continuous blue curve in Figure~\ref{fig:mcp-adiabatic}(c). The field in which the $| 70\mathrm{c_+}\rangle$ state ionizes is indicated in this figure by the right-most purple shaded vertical band. To confirm that this evolution of the $| 70\mathrm{c'_+}\rangle$ state into the $| 70\mathrm{c_+}\rangle$ circular state in the pure magnetic field occurs adiabatically, the electric field was also switched off more rapidly, e.g., in a time $\Delta t_{\mathrm{off}} = 20$~ns. As in Figure~\ref{fig:mcp-adiabatic}(b), this strongly modified the electron signal observed. However for switching times greater than $\sim~500$~ns the detected signal did not change significantly, confirming adiabatic evolution to the $| 70\mathrm{c_+}\rangle$ state. As in the case of the initially laser photoexcited $|73\mathrm{s}\rangle$ state, the electric field ionization signal corresponding to the $| 70\mathrm{c_+}\rangle$ state exhibits both adiabatic and non-adiabatic pathways to ionisation. These pathways differ from those followed in the ionization of the $|73\mathrm{s}\rangle$ state and therefore make it difficult to precisely quantify the efficiency of transferring atoms to the target $| 70\mathrm{c_+}\rangle$ state. However, from the normalized amplitude of the dominant feature at 39.4~V/cm in Figure~\ref{fig:mcp-adiabatic}(c), which corresponds to the field ionization of the $| 70\mathrm{c_+}\rangle$ state, this efficiency was estimated to be $\sim80$~\%.
	
	\section{Adiabatic dynamics}
	
	The rate at which ensembles of atoms in circular Rydberg states can be prepared using this modified crossed-fields method is limited by the times required for adiabatic polarization of the laser photoexcited low-$\ell$ Rydberg state and adiabatic de-polarization of the $| n\mathrm{c'_+}\rangle$ state. If implementing this method for the preparation of circular states with higher values of $n$, or in other systems, the criteria to ensure adiabatic evolution of these states must be considered. The probability of adiabatic evolution through an avoided crossing involving two isolated levels, $| \mathrm{a} \rangle$ and $| \mathrm{b} \rangle$, with energies that change linearly in time in the diabatic basis, can be determined using Landau-Zener theory \cite{Zener1932NonAdiabaticLevels}, where the probability of adiabatic passage is,
	\begin{eqnarray}
	P_{\mathrm{adiabatic}} & = & 1 - \exp \left( -\frac{2 \pi |V_{ij}|^2}{\hbar \, (\mathrm{d}E/\mathrm{d}t)} \right) \\
	\label{eq:landau-zener}
	& = & 1 - \exp \left( -\frac{2 \pi |V_{ij}|^2}{\hbar \, (\mathrm{d}E/\mathrm{d}F)(\mathrm{d}F/\mathrm{d}t)} \right)
	\end{eqnarray}
	with $E$ the energy difference between the levels, $F$ the electric field strength, and $V_{ij} = E_{\mathrm{min}}/2$ with $E_{\mathrm{min}}$ the minimum energy interval between the two states.
	
\begin{figure}
		\includegraphics[width=0.5\textwidth]{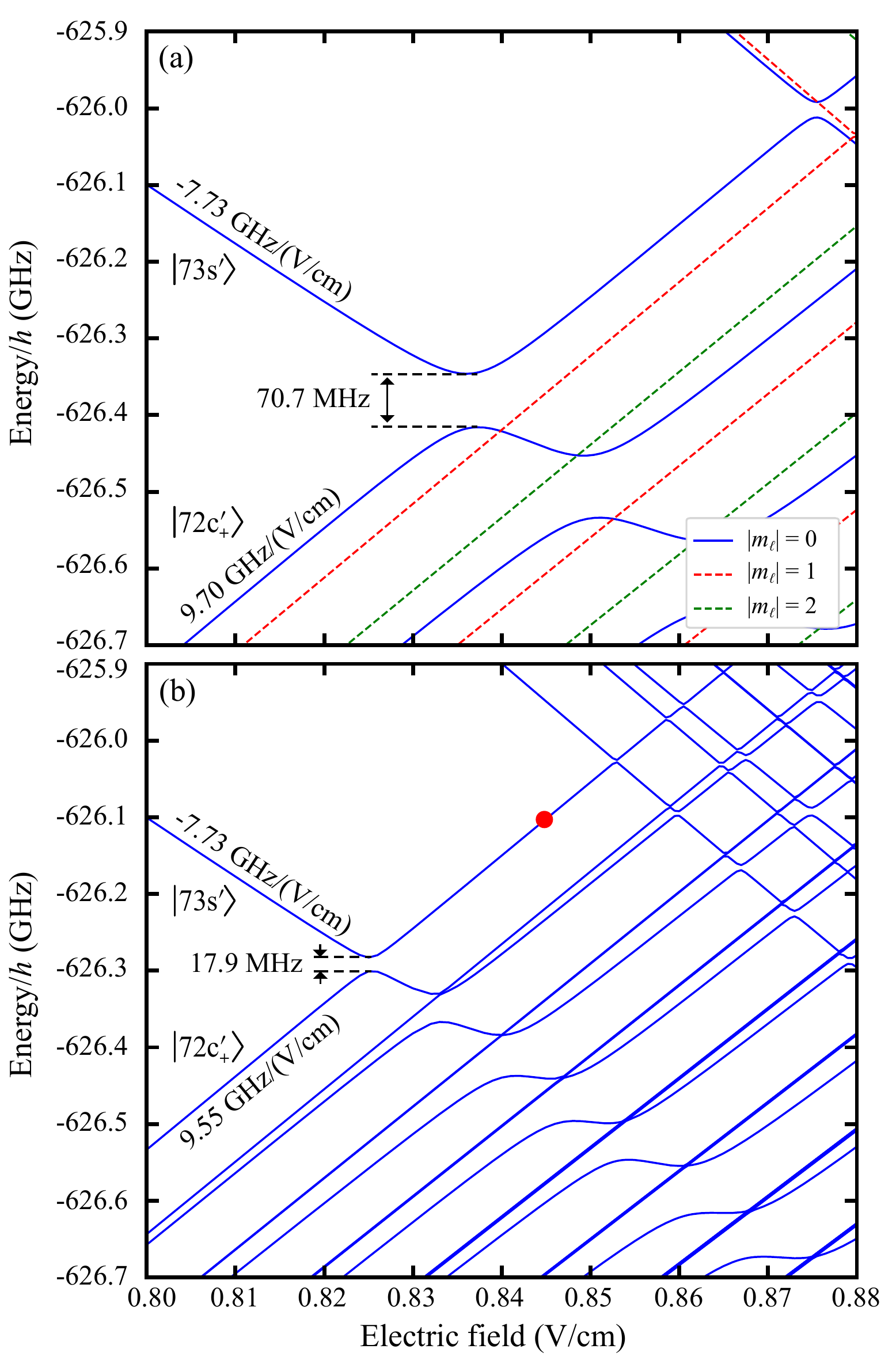}
		\caption{\label{fig:stark-map-comparison} Calculated energy-level structure of triplet Rydberg states of helium in a magnetic field of (a) $B_z = 0$~G, and (b) $B_z = 16.154$~G, and a perpendicular electric field. The red circle indicates the position from which microwave transfer to the $| 70\mathrm{c'_+}\rangle$ state was carried out after adiabatic passage through the avoided crossing between the $| 73\mathrm{s'} \rangle$ and $| 72\mathrm{c'_+}\rangle$ states.}
	\end{figure}

	In Figure~\ref{fig:stark-map-comparison}(b), an expanded view of the Stark map in Figure~\ref{fig:stark-map-full} is displayed which highlights the energy level structure in the vicinity of the first avoided crossing encountered by the $| 73\mathrm{s'} \rangle$ state. In a pure electric field, Figure~\ref{fig:stark-map-comparison}(a), i.e., when $B_{z}=0$, $m_{\ell}$ is a good quantum number and the interacting levels are only those which have the same values of $m_{\ell}$. For the $| 73\mathrm{s'} \rangle$ state and the $n=70$ Stark state with which it interacts, this leads to a minimum energy interval between the levels ot the avoided crossing corresponding to 70.7~MHz and electric dipole moments of the two states in the diabatic basis of 15355~D [-7.73~GHz/(V/cm)] and $-19268$~D [9.70~GHz/(V/cm)], respectively. As pointed out recently by Liu et al. \cite{Liu2018SymmetryBreaking}, the addition of the perpendicular magnetic field, in this case a field of $B_{z}=16.154$~G, breaks the symmetry of the Hamiltonian [Equation~(\ref{eq:hamiltonian-zeeman-stark})], with the result that the corresponding avoided crossing reduces in size to 17.9~MHz. This effect can be seen in Figure~\ref{fig:stark-map-comparison}(b). In this case, encountered in the experiments described here, the smaller avoided crossing is a consequence of the weaker coupling between the $| 73\mathrm{s'} \rangle$  and $| 72\mathrm{c'_+}\rangle$ states by the electric field because of $m_{\ell}$-mixing and additional interactions with the outer Stark states of the $n=72$ manifold. Therefore the conditions for adiabatic traversal of this avoided crossing in the presence of the perpendicular magnetic field are stricter than in a pure electric field.
	
\begin{figure}
		\includegraphics[width=0.5\textwidth]{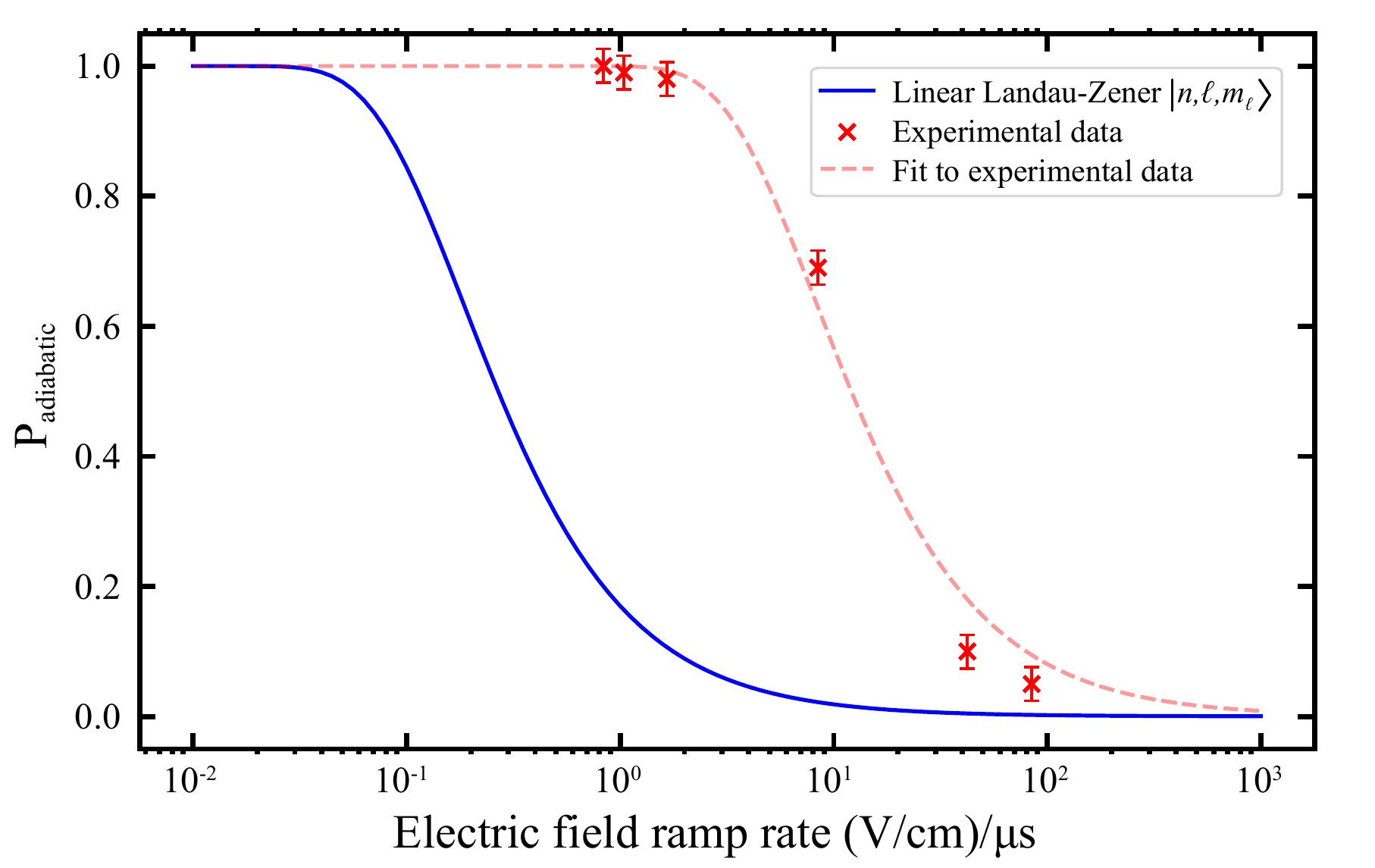}
		\caption{\label{fig:LZ-comparison} Probabilities of adiabatic passage through the avoided crossing between the $| 73\mathrm{s'} \rangle$ and $| 72\mathrm{c'_+}\rangle$ states in a magnetic field of 16.154~G. The experimental data (red crosses), and results of the calculations in the $|n, \ell, m_{\ell} \rangle$ basis (continuous blue curve) are shown.}
	\end{figure}

	From the data shown in Figure~\ref{fig:mcp-adiabatic}(b), and additional experiments performed for switch-off times, $\Delta t_{\mathrm{off}}$, of 1010~ns, 510~ns, 100~ns, and 10~ns, probabilities for adiabatic traversal of the avoided crossing in Figure~\ref{fig:stark-map-comparison}(b) were determined from the amplitude of the electron signals. These probabilities, displayed with respect to the corresponding rate of change of the applied electric field, are shown as red crosses in Figure~\ref{fig:LZ-comparison}. A function with the general form of that associated with the linear Landau-Zener probability in Equation~(\ref{eq:landau-zener}), fit to the data and displayed as the dashed red curve, demonstrates that the dynamics at the avoided crossing change exponentially from adiabatic to diabatic as the rate of change of the electric field is increased. The convergence of the measured data for the slowest rates, i.e., those close to 1~(V/cm)/$\mu$s, demonstrates that adiabatic evolution is achieved in these cases. However, upon comparison of the measured data with the results obtained using the linear Landau-Zener model and the calculated parameters associated with the avoided crossing in Figure~\ref{fig:stark-map-comparison}(b) (continuous blue curve in Figure~\ref{fig:LZ-comparison}), a significant difference is observed. This behavior can result from deviations of the characteristics of this avoided crossing from those of an ideal two-level system with linear Stark energy shifts in the diabatic basis, or arise because the Hamiltonian in Equation~(\ref{eq:hamiltonian-zeeman-stark_general}) does not provide a sufficiently complete description of the atom under the experimental conditions. In order that the linear Landau-Zener model accurately describe the experimental data in Figure~\ref{fig:LZ-comparison}, it would be necessary that the avoided crossing be 6.7 times the calculated value in Figure~\ref{fig:stark-map-comparison}(b). Such a discrepancy (on the order of $100$~MHz) between the experimental data and the calculated energy level structure is not observed in Figure~\ref{fig:transition-spectra}.
	
	As discussed in Section~III, for the weak magnetic fields employed in the experiments described here, the inclusion of the diamagnetic interaction in the calculations results in relative shifts of the $| 73\mathrm{s'} \rangle$ and $| 72\mathrm{c'_+}\rangle$ states of $<1.0$~MHz. Corrections on this scale do not significantly modify the Landau-Zener dynamics at this crossing. On the other hand, contributions from the spin-orbit interaction in these triplet Rydberg states can have a more pronounced effect. To characterize the effects of the spin-orbit interaction in the experimental data, calculations were performed in the $|n, L, S, J, M_{J} \rangle$ basis. Because of computational limitations only the larger, diagonal spin-orbit matrix elements, and hence the basis states with $S=1$, were considered in these calculations. Under these conditions, including the same range of values of $n$ as in the calculation of the data shown in Figure~\ref{fig:stark-map-full} would have resulted in a basis size of 92073 which is beyond our computational capabilities. Therefore with the aim of making a qualitative comparison of the characteristics of the avoided crossing between the $| 73\mathrm{s'} \rangle$ and $| 72\mathrm{c'_+}\rangle$ states, with and without the effects of the spin-orbit interaction, the energy level structure at the crossing was determined in the $|n, L, S, J, M_{J} \rangle$ and $|n, \ell, m_{\ell} \rangle$ bases containing states for which $72 \leq n \leq 73$, respectively. The size of these bases are 31539 and 10513, respectively.
	
	\begin{figure}
		\includegraphics[width=0.5\textwidth]{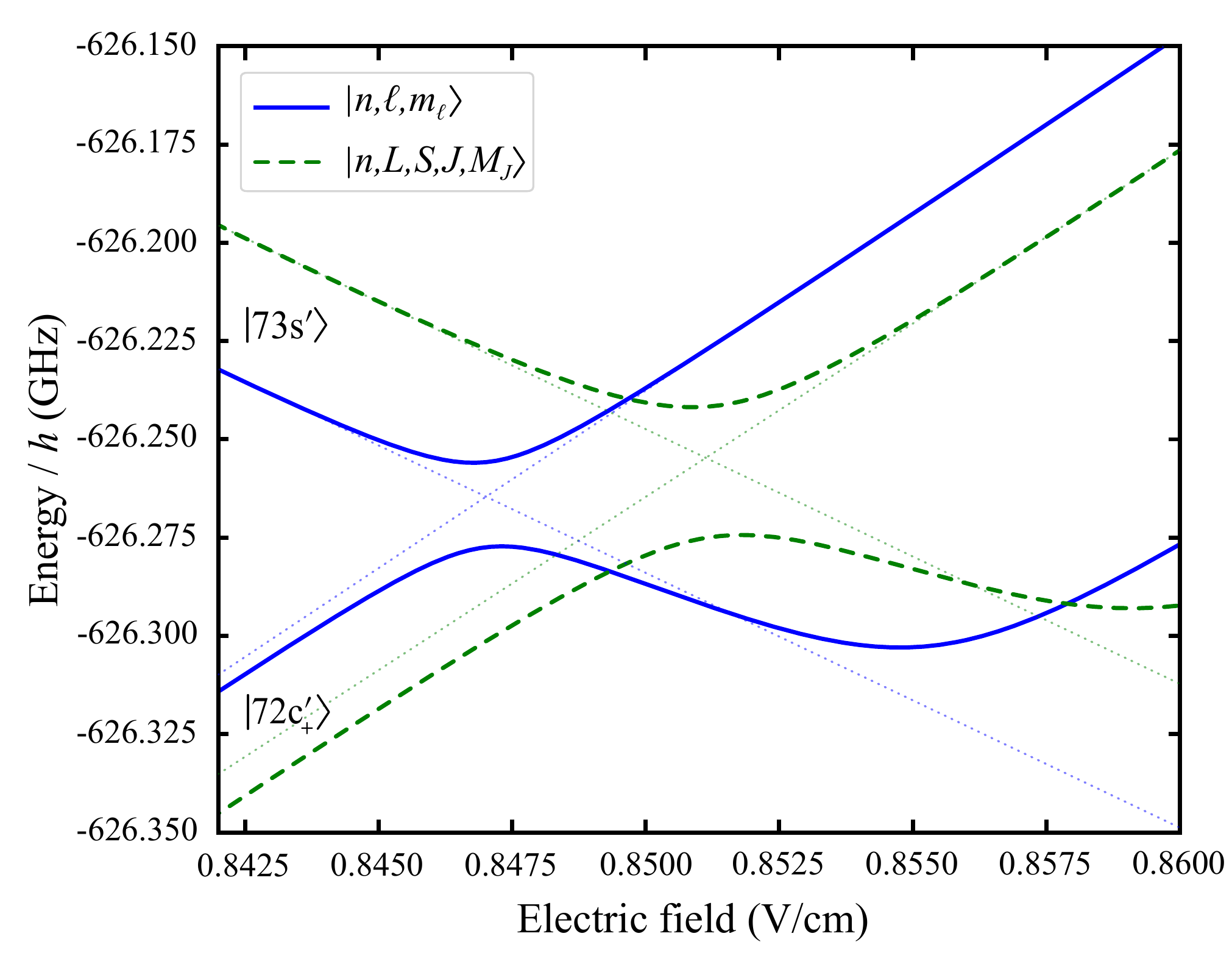}
		\caption{\label{fig:basis_comparison} Avoided crossing between the $| 73\mathrm{s'} \rangle$ and $| 72\mathrm{c'} \rangle$ states, calculated in the $|n, \ell, m_{\ell} \rangle$ (continuous blue curve) and $|n, L, S, J, M_{J} \rangle$ (dashed green curve) bases containing $S=1$ levels with $72 \leq n \leq 73$. }
	\end{figure}
	
The results of these calculations are displayed in Figure~\ref{fig:basis_comparison}.	From the data in this figure, the minimum interval between the $| 73\mathrm{s'} \rangle$ and $| 72\mathrm{c'_+}\rangle$ states in Figure~\ref{fig:basis_comparison} corresponds to 21.64~MHz and 33.26~MHz, in the $|n, \ell, m_{\ell} \rangle$ and $|n, L, S, J, M_{J} \rangle$ bases, respectively. The value obtained from the calculations in the $|n, \ell, m_{\ell} \rangle$ basis is larger than that in Figure~\ref{fig:stark-map-comparison}(b), and the crossing occurs at a higher field than that in Figure~\ref{fig:stark-map-comparison}(b) because of the reduced basis size. In this smaller basis the calculations have not fully converged, however they do allow for a direct comparison to be made with the calculations in the $|n, L, S, J, M_{J} \rangle$ basis. The ratio of the minimum interval between the $| 73\mathrm{s'} \rangle$ and $| 72\mathrm{c'_+}\rangle$ states in the $|n, \ell, m_{\ell} \rangle$ and $|n, L, S, J, M_{J} \rangle$ bases in Figure~\ref{fig:basis_comparison} is 1.54. Using this ratio, it may be expected that the true size of the avoided crossing in the $|n, L, S, J, M_{J} \rangle$ basis can be determined by scaling the value obtained in the larger $|n, \ell, m_{\ell} \rangle$ basis after convergence [i.e., Figure~\ref{fig:stark-map-comparison}(b)] by this same factor. Following this reasoning, the minimum interval between the $| 73\mathrm{s'} \rangle$ and $| 72\mathrm{c'_+}\rangle$ states in the $|n, L, S, J, M_{J} \rangle$ basis is expected to be $17.9$~MHz~$ \times 1.54=27.6$~MHz, approximately 10~MHz larger than that in Figure~\ref{fig:stark-map-comparison}(b) where the spin-orbit interaction was neglected. This accounts for some, but not all, of the discrepancy between the linear Landau-Zener model and the experimental data.
	
	The linear Landau-Zener model assumes an ideal two-level system, in which the gradients, i.e., the static electric dipole moments, of the states on either side of the avoided crossing are constant in the diabatic basis. The avoided crossings shown in Figure~\ref{fig:basis_comparison} do not fulfill this criteria. This is emphasized by the dotted lines fit to the $| 73\mathrm{s'} \rangle$ state in the adiabatic basis before and after the avoided crossing in each case. Using these lines as a guide, it can be seen that (1) the gradient of the $| 72\mathrm{c'_+}\rangle$ state on the high-field side of the crossing is smaller than that of the $| 73\mathrm{s'} \rangle$ state on the low-field side, (2) this gradient is not constant within the range of electric fields in Figure~\ref{fig:basis_comparison}, (3) the electric field strength at which the turning points of the adiabatic states, i.e., the points at which the electric dipole moments are zero, is different for the $| 73\mathrm{s'} \rangle$ and $| 72\mathrm{c'_+}\rangle$ states, and (4) there is a displacement between the adiabatic states before and after the avoided crossing, as indicated by the dotted lines in Figure~\ref{fig:basis_comparison}. Together, these features increase the adiabadicity of the passage through the avoided crossing for an applied time-dependent field compared to that expected from the linear Landau-Zener model. Further, these features are more pronounced in the calculations in the $|n, L, S, J, M_{J} \rangle$ basis, suggesting that the spin-orbit interaction must be accounted for when accurately determining the probabilities for adiabatic transfer.
	
	It has been shown in Ref.~\cite{JoanLasinio2003AsymmetricLandauZener} that an avoided crossing between two energy levels with linear and quadratic dependencies, respectively, can lead to deviations from the linear Landau-Zener model, but maintains the exponential transition from adiabatic to diabatic passage. In order to capture the dynamics of the avoided crossings shown in Figure~\ref{fig:basis_comparison}, it would be necessary to extend this model to include all of the features discussed above. Because of the challenges in accurately calculating the atomic energy level structure in crossed electric and magnetic fields in the $|n, L, S, J, M_{J} \rangle$ basis at the values of $n$ employed in the experiments reported here, which are further exacerbated at higher values of $n$, we conclude that in implementing this modified crossed-field method for the preparation of high-$n$ circular states, calculations in the $|n, \ell, m_{\ell} \rangle$ basis can only be used as a guide when determining the parameters for adiabatic passage through such avoided crossings. Accurate values must therefore ultimately be determined experimentally as has been demonstrated here. This problem however may motivate future theoretical investigations, particularly in the context of effects of symmetry breaking on the dynamics of quantum systems.
	
	% ------------------------------------------------
	%             Magnetic field considerations
	% ------------------------------------------------
	
	\section{\label{sec:discussion}Range of applicability}
	
	The modified crossed-fields method presented here can be applied to the preparation of circular Rydberg states in a wide range of atoms and molecules with minimal restrictions on the spectral resolution required in the initial laser photoexcitation step. In addition to being of importance in the preparation of circular Rydberg states with high values of $n$, this aspect of the method is advantageous in general in the preparation of circular Rydberg states in species for which laser photoexcitation must be carried out using pulsed laser radiation in the ultraviolet or vacuum ultraviolet regions of the electromagnetic spectrum where high spectral resolution is difficult to achieve because of laser pulse length limitations and Doppler effects. The primary requirements for the implementation of this circular state preparation scheme are that appropriate low-$\ell$ states, i.e., s or p Rydberg states, have (1) sufficiently large quantum defects that they can be resolved from the manifold of degenerate higher-$\ell$ states and therefore selectively photoexcited, (2) sufficiently long lifetimes, $\gtrsim 100$~ns, and (3) sufficiently large avoided crossings at the Inglis-Teller limit that they can be adiabatically polarized to acquire similar static electric dipole moments to the target circular states in the same electric fields. 
	
	The minimum time required to implement circular state preparation at a particular value of $n$ using this scheme is ultimately set by the strength of the magnetic field employed. The upper limit on the magnitude of this field is imposed by the quantum defect of the initially excited low-$\ell$ state, and its effect on the size of the avoided crossing encountered by this state at the Inglis-Teller limit. Because of its dependence on the linear Zeeman splitting between $m_{\ell}$ sublevels in zero electric field, adiabatic evolution from the $| n \mathrm{c'_+}\rangle$ to $| n \mathrm{c_+}\rangle$ state can be implemented most rapidly in the largest possible magnetic field. In the case of interest here -- the preparation of triplet $| n \mathrm{c_+}\rangle$ states in helium -- the magnetic field strength for which the $| 73 \mathrm{c_-} \rangle$ state crosses the energy of the initially excited $| 73 \mathrm{s} \rangle$ state in a zero electric field is $\sim 48$~G. This therefore sets a first bound on the magnetic field that can be considered in implementing this scheme. Symmetry breaking in the crossed electric and magnetic fields close to the Inglis-Teller limit results in a reduction in the size of the avoided crossing between the $| 73 \mathrm{s'} \rangle$ and $| 72 \mathrm{c'_+}\rangle$ states as the magnetic field strength is increased. For example, this crossing exhibits a minimum interval corresponding to 70.7~MHz for $B_{z}=0$~G, which reduces to 17.9~MHz ($\sim 27.6$~MHz if spin-orbit contributions are accounted for, see Section~VI) for $B_{z}=16.154$~G. This smaller avoided crossing must be traversed more slowly to ensure adiabatic evolution. Based on the discussion in Section~VI, optimization of the magnetic field strength to engineer this avoided crossing such that it allows for more rapid circular state preparation must be done experimentally. In the experiments described here the linear Zeeman splitting between the $m_{\ell}$ sublevels with $n=70$ in zero electric field corresponds to 22.6~MHz. Therefore these are of a similar order of magnitude as the 17.9~MHz avoided crossing between the $| 73 \mathrm{s'} \rangle$ and $| 72 \mathrm{c'_+}\rangle$ states. The smaller static electric dipole moment of the $| 70 \mathrm{c'_+}\rangle$ state near zero electric field than those of the $| 73 \mathrm{s'} \rangle$ and $| 72 \mathrm{c'_+}\rangle$ states at their avoided crossing ensures that a similar switching time of $\sim 1$~$\mu$s leads to adiabatic evolution.
	
	This method can be extended to the preparation of high-$n$ circular Rydberg states in other atoms and in molecules. For example, such states with values of $n>100$, in alkali metal atoms are of interest in studies of Rydberg-electron interactions with ultracold gases~\cite{Balewski2013CouplingSingleElectron}. In these settings new opportunities for imaging bound-state electron probability density distributions are also expected~\cite{Karpiuk2015ImagingSingleRydberg}. Circular Rydberg states with high values of $n$ are also of interest in studies of ion-neutral reactions at low translational temperatures where the Rydberg electron can be exploited to shield the reaction center from stray electric fields and hence ensure that low collision energies are maintained throughout the reaction process~\cite{Allmendinger2016H2LowCollision,Schlagmuller2016UltracoldReactions}. In the case of the $\mathrm{H}_2^{+} + \mathrm{H}_2$ reaction, studied at low temperature through collisions of Rydberg H$_2$ with ground-state H$_2$~\cite{Allmendinger2016H2LowCollision}, the method described here may be applied to the preparation of circular Rydberg states of H$_2$ to open new opportunities for investigating, and if necessary minimizing, the contributions from the Rydberg electron to the reaction dynamics. 
	
	% ------------------------------------------------
	%                   Applications
	% ------------------------------------------------
	
	\section{\label{sec:applications}Applications}
	
\begin{figure}
		\includegraphics[width=0.5\textwidth]{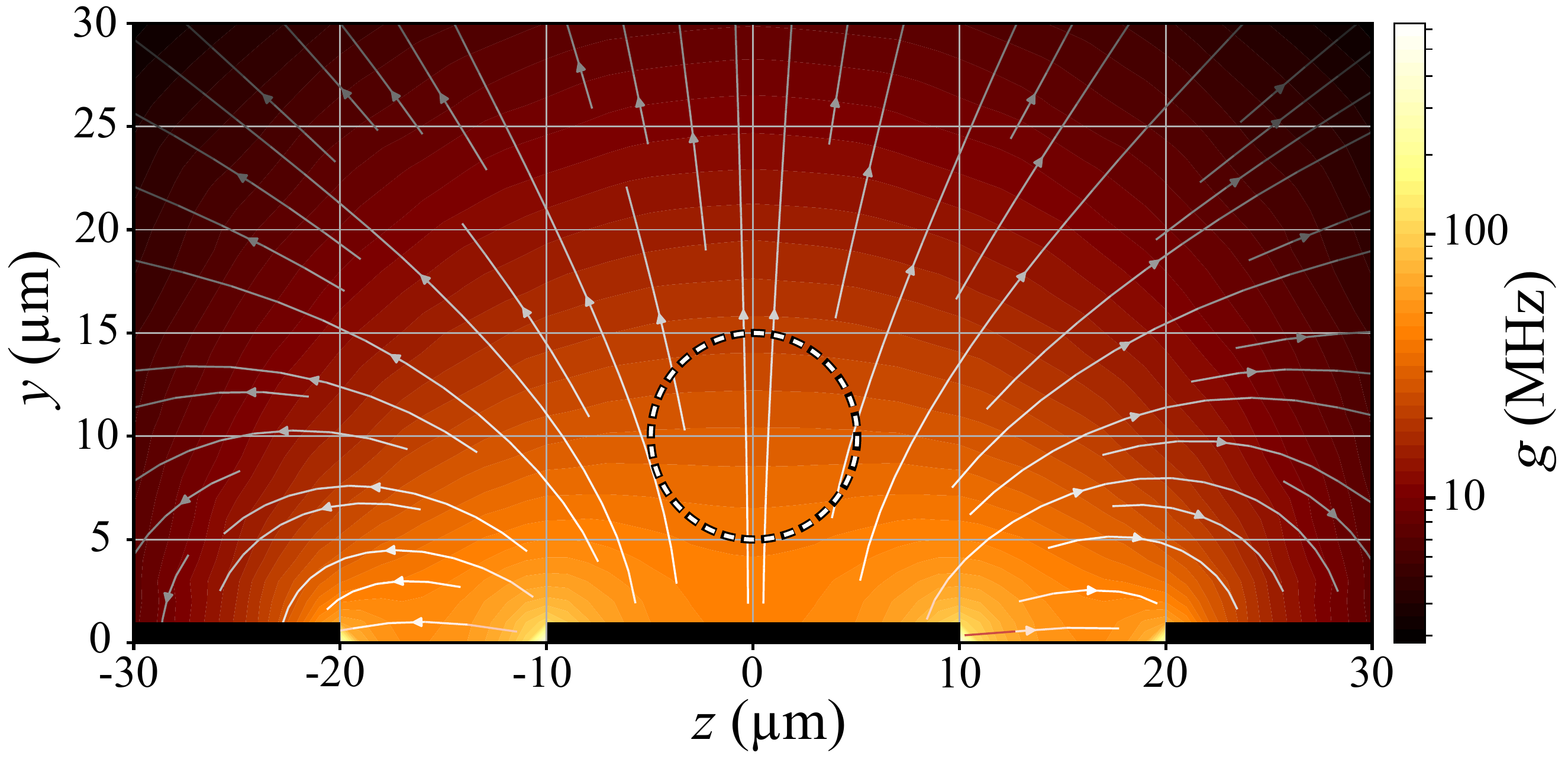}
		\caption{\label{fig:g-map} Coupling strength, $g$, between the 18.777~GHz vacuum field of a two-dimensional superconducting microwave resonator and the $|70\mathrm{c_+}\rangle$~$\rightarrow$~$|71\mathrm{c_+}\rangle$ circular-to-circular Rydberg state transition in helium.}
	\end{figure}

	The helium atoms in the $| 70\mathrm{c_+}\rangle$ state prepared in the experiments described here are ideally suited to applications in hybrid cavity QED, whereby they are coupled to two-dimensional superconducting microwave resonators. In such experiments, an important parameter to optimize is the atom-resonator coupling strength above the chip surface. By way of example, a map of the vacuum coupling strength, $g$, for the $|70\mathrm{c_+}\rangle \rightarrow |71\mathrm{c_+}\rangle$ circular-to-circular Rydberg state transition in helium and a superconducting co-planar waveguide resonator is shown in Figure~\ref{fig:g-map}. The corresponding microwave field distribution was calculated using finite-element methods and normalized for the vacuum field of the first harmonic of the resonator. In this calculation the center conductor was 20~$\mu$m wide, with 10~$\mu$m gaps separating it from the ground planes on either side. The superconducting layer was 100~nm thick, and was supported on top of a 525~$\mu$m thick, undoped, high resistivity ($> 8000$~$\Omega\,$cm) silicon substrate with a relative permittivity of $\epsilon_{\mathrm{r}}=11.68$. The $|70\mathrm{c_+}\rangle \rightarrow |71\mathrm{c_+}\rangle$ transition in helium occurs at a frequency of 18.777~GHz in the absence of electric and magnetic fields and has an electric dipole transition moment of $3477 \, e a_0$. The $|70\mathrm{c_+}\rangle$ and $|71\mathrm{c_+}\rangle$ states have radiative lifetimes of 155~ms and 166~ms, respectively. In this case, the amplitude of the vacuum field 10~$\mu$m above the center conductor is 0.104~V/cm, leading to a coupling strength, at this position, of 29.0~MHz. 
	
	Stray electric fields from patch potentials across the surface or charged particles adsorbed in the insulating gaps can lead to variations in atomic transition frequencies, electric dipole transition moments and static electric dipole moments across a cloud of atoms. This can present a range of challenges to achieving strong atom-resonator coupling including variations in the detuning between resonance frequencies of the atoms and the resonator. This could complicate the atom-resonator coupling and reduce the coherence across the cloud of atoms. It is therefore important to consider the range of electric fields that might be expected in such a setting.
	
	The work presented in Ref.~\cite{Carter2011EnergyShifts} described a model from which the root-mean-square, $F_{\mathrm{rms}}$, and variance, $\sigma^2$, of the distribution of stray electric fields emanating from a surface exhibiting patch potentials could be calculated. Using the experimental data in Ref.~\cite{Pu2010IonizationRydberg}, the patch potentials across an atomically flat Au surface, $F_{\mathrm{rms}}$ at distances of 5~$\mu$m, 10~$\mu$m, and 15~$\mu$m was extrapolated to give 201~mV/cm, 51~mV/cm, and 22~mV/cm, respectively. The standard deviation, $\sigma$, of the electric field in these cases was $0.72 \, F_{\mathrm{rms}}$. For a cloud of atoms with a diameter of 10~$\mu$m and centered 10~$\mu$m above the center conductor (white/black dashed circle in Figure~\ref{fig:g-map}), the values of $F_{\mathrm{rms}}$ cover a range of 179~mV/cm. This represents a lower limit on the magnitude and range of the stray electric fields expected above a superconducting chip such as in Figure~\ref{fig:g-map}.
	
	An analysis of stray electric fields generated by charges adsorbed onto superconducting surfaces and in the insulating gaps of co-planar waveguide structures has been carried out in Ref.~\cite{Thiele2015ImagingElectric}. For a co-planar waveguide with center conductor and gap widths of 180~$\mu$m and 80~$\mu$m, respectively, charge densities of $\sigma_{\mathrm{s}} = -2.10(5) \times 10^{-6}$~C/m$^2$ and $\sigma_{\mathrm{g}} = -23.6(1) \times 10^{-6}$~C/m$^2$ were deduced for the surface and insulating gaps, respectively. For the geometry of the device in Figure~\ref{fig:g-map}, the electric fields generated by a charge density of $\sigma_{\mathrm{g}} = -23.6(1) \times 10^{-6}$~C/m$^2$ in the insulating gaps have been calculated using finite element methods. In this case at distances of 5~$\mu$m, 10~$\mu$m, and 15~$\mu$m above the surface, electric fields of 189~V/cm, 181~V/cm, and 172~V/cm, respectively, arise with an average gradient of 1.65~kV/cm$^2$. These values are approximately a factor of 2 smaller than those calculated using the co-planar waveguide dimensions used in Ref.~\cite{Thiele2015ImagingElectric}. It is possible to further reduce stray electric fields by compensating them through the application of appropriate potentials to surrounding electrodes \cite{Thiele2014ManipulatingRydberg}.
	
	To obtain a decoherence time, $T_2$, of 10~$\mu$s for an ensemble of atoms coupled to a resonator mode on a chip exhibiting stray electric fields, it is necessary that the range of transition frequencies across the ensemble, due to variations in the electric field, is less than 100~kHz. To maximise the magnitude of the stray electric fields which can be tolerated to achieve this coherence time, it is therefore desirable to minimise the sensitivity of the circular-to-circular state transition frequencies to electric fields. Although these transitions are already less sensitive than those of low-$\ell$ states, their sensitivity can be further reduced by minimizing the magnetic field because this removes the linear Stark shift, leaving the quadratic Stark shift as the dominant term. For the $|70\mathrm{c_+}\rangle$~$\rightarrow$~$|71\mathrm{c_+}\rangle$ transition, in the absence of a magnetic field, electric field distributions which correspond to a range of 100~kHz in the transition frequency are, $0 \pm 274$~mV/cm, $100 \pm 137$~mV/cm, and $200 \pm 94$mV/cm. It is possible to further reduce the sensitivity of the atoms to electric fields through the use of non-resonant dressing fields to null the differential polarizability of consecutive circular states, as has recently been proposed in Ref.~\cite{Ni2015NonresonantDressing}, leaving the quartic Stark shift as the dominant term.
	
	% ------------------------------------------------
	%                   Conclusion
	% ------------------------------------------------
	
	\section{\label{sec:conclusion}Conclusion}
	
	We have presented a modified version of the crossed-fields method which allows the selective preparation of high-$n$ circular Rydberg states in atoms and molecules. By implementing a microwave transfer step from an initially excited, and subsequently strongly polarized, low-$\ell$ state to the target circular state with a similar static electric dipole moment, the sensitivity of this circular state preparation scheme to DC electric field inhomogeneities and low-frequency electric field noise is significantly reduced compared to the standard crossed-fields method. The experimental results presented here demonstrate the implementation of this method for the preparation of the $n=70$, $m_{\ell}=+69$ circular Rydberg state in helium. In this case the efficiency of converting atoms from the initially laser photoexcited $| 73\mathrm{s} \rangle$ state to this circular state was $\sim$80\%. This could be increased by improving the coherence of the microwave transfer by performing the experiment in a single mode microwave waveguide. 
	
	The long radiative lifetimes of these high-$n$ circular Rydberg states, their low sensitivity to stray electric fields, resulting long coherence times, and their large electric dipole transition moments make them ideally suited for interfacing with superconducting circuits. The selective preparation of such states is an essential capability for future hybrid quantum systems involving Rydberg atoms.
	
	\begin{acknowledgments}
		This work was supported by the Engineering and Physical Sciences Research Council under Grant No. EP/L019620/1 and through the EPSRC Centre for
		Doctoral Training in Delivering Quantum Technologies.
		
		The authors acknowledge the use of the UCL Legion High Performance Computing Facility (Legion@UCL), and associated support services, in the completion of this work.
	\end{acknowledgments}
	
	\bibliography{citations}
	
\end{document}